\documentclass[10pt,journal,twoside]{IEEEtran}

\usepackage{cite}
\usepackage{epsfig}
\usepackage{epstopdf}
\usepackage{graphicx}
\usepackage{amsfonts}
\usepackage{amsmath}
\usepackage{amssymb}
\usepackage{multicol}
\usepackage{multirow}
\usepackage{psfrag}
\usepackage{subfigure}
\usepackage{stfloats}
\usepackage{footnote}
\usepackage{array}
\usepackage{algorithm}
\usepackage{algpseudocode}
\usepackage{color}

\IEEEoverridecommandlockouts

\let\ss= \scriptscriptstyle
\DeclareMathOperator*{\E}{\mathbb{E}}

\begin{document}

\title{A Survey on Estimation Schemes in Molecular Communications}

\author{Xinyu Huang, Yuting Fang, and Nan Yang
\thanks{X. Huang and N. Yang are with the School of Engineering, Australian National University, Canberra, ACT 2600, Australia (email: \{xinyu.huang1, nan.yang\}@anu.edu.au).}
\thanks{Y. Fang is with the Department of Electrical and Electronic Engineering, University of Melbourne, Parkville, VIC 2010, Australia (e-mail: yuting.fang@unimelb.edu.au).}}

\markboth{Submitted to Digital Signal Processing}{Huang \MakeLowercase{\textit{et al.}}: A Survey on Estimation Schemes in Molecular Communications}

\maketitle

\begin{abstract}
This survey paper focuses on the estimation schemes in molecular communication (MC) systems. The existing studies in estimation schemes can be divided into parameter estimation (e.g., distance, diffusion coefficient, and flow velocity) and channel estimation.  In this paper, we present, for the first time, a comprehensive survey on i) distance estimation, since distance is the most widely estimated parameter in current studies, ii) estimation of other parameters (i.e. the parameters excluding distance), and iii) channel estimation that focuses on the channel impulse response (CIR). Moreover, we examine the noise that may impact on the estimation performance and the metrics applied to evaluate the performance of different estimation schemes. Numerical results are provided to compare the performance of different distance estimation schemes. In addition, future research directions in parameter estimation and channel estimation are identified and discussed.
\end{abstract}

\begin{IEEEkeywords}
Molecular communication, parameter estimation, channel estimation, statistical model.
\end{IEEEkeywords}

\section{Introduction}

Molecular communications (MC) is an emerging technology in the past decade, which has great potential to facilitate nano-scale communication. Instead of using electromagnetic waves as information carriers as in traditional wireless communication, MC uses small particles such as molecules or lipid vesicles to deliver information \cite{farsad2016comprehensive}. Moreover, MC owns unique merits such as biocompatibility and low energy consumption, which make MC more suitable for \textit{in vivo} applications than other communication methods, e.g., electromagnetic methods.

Recently, some surveys and tutorials have discussed the benefits and challenges of MC from different perspectives, e.g., \cite{farsad2016comprehensive,chude2017molecular,jamali2019channel,kuscu2019transmitter,nakano2019methods}. Specifically, \cite{farsad2016comprehensive} presented a detailed introduction on MC and provided an overall survey at the recent advances in the micro-scale MC and the macro-scale MC. In \cite{chude2017molecular}, the authors presented a survey on the applications of MC and molecular networks with the focus on targeted drug delivery. In \cite{jamali2019channel}, Jamali et al. provided a tutorial review on mathematical channel modeling for diffusive MC systems. In \cite{kuscu2019transmitter}, the authors reviewed the contributions to the architectures of transmitter $(\mathrm{TX})$ and receiver $(\mathrm{RX})$ among nanomaterial-based nanomachines and/or biological entities and provided a complete overview of modulation, coding, and detection techniques employed for MC. Nakano et al. in \cite{nakano2019methods} provided a comprehensive review on mobile MC. Although these studies stand on their own merits, the estimation schemes in MC have yet to be reviewed and summarized.

The estimation schemes investigated in current studies can be classified into two categories, namely, parameter estimation and channel estimation. Estimated parameters usually include the distance between the $\mathrm{TX}$ and the $\mathrm{RX}$, the diffusion coefficient of molecules, the flow velocity in the MC environment, and so on. Among these parameters, distance estimation is the most popular research area in current studies. This is because the distance between nanomachines is one of the most pivotal parameters for the communication channel. Specifically, we summarize the significance of distance estimation in MC systems as follows:
\begin{itemize}
\item Distance estimation can be utilized to improve the channel performance since the distance affects the transmission rate. If a $\mathrm{TX}$ obtains the knowledge about the distance, it can adjust the number of released molecules to achieve a high probability of molecules arriving at a $\mathrm{RX}$ and avoid using too many molecules to reduce interferences, such as inter-symbol interference (ISI) and inter-link interference (ILI).
\item In the application of targeted drug delivery \cite{chude2017molecular}, it is highly important to know the accurate location of the target site, e.g., a tumor, in the human body such that drugs are delivered to this site. The target site can be localized by estimating the relative distance between the tumor and nanomachines \cite{moore2011addressing}.
\end{itemize}
It is noted that in biological systems, there exist some techniques to determine the distance between two nanomachines in MC. For example, a cell can estimate the relative distance from an organism via producing a type of molecules to establish a chemical gradient \cite{tostevin2007fundamental}.

Apart from distance estimation, estimating other parameters (i.e., the parameters excluding the distance) is also essential for some promising MC applications. For example, the MC system can be deployed in a blood-vessel environment for the healthcare application. In this application, estimating the blood flow velocity can help to measure the blood pressure. Also, estimating the diffusion coefficient of molecules can help to determine the blood composition and identify major changes in blood cell counts \cite{ho2004white}. Furthermore, estimating the degradation rate of molecules can help to measure the pH level of blood since molecule degradation varies with the pH level \cite[Ch. 10]{chang2005physical}. Motivated by these benefits, some studies, e.g., \cite{moore2012measuring,wang2015distance,noel2015joint}, have proposed different methods for parameter estimation. Specifically, the estimated parameters include the number of emitted molecules, the degradation rate of molecules, the flow velocity, the diffusion coefficient of molecules, the release time of molecules, the clock offset between the $\mathrm{TX}$ and the $\mathrm{RX}$, the start time of each symbol interval, and the signal-to-noise ratio (SNR).

Different from parameter estimation, channel estimation focuses on estimating the channel impulse response (CIR) in MC systems. Here, the CIR is defined as the probability of observing one molecule at the $\mathrm{RX}$ at time $t$ when molecules are impulsively released at time $t_0=0$. The CIR is important for the design of equalization and detection schemes in MC systems \cite{noel2014optimal,mahfuz2014comprehensive}. Motivated by this importance, a few studies have investigated the estimation of the CIR via different methods, e.g., \cite{7546910,8685177}.

In this paper, we divide estimation schemes into three categories: Distance estimation, estimation of other parameters (i.e., the parameters excluding the distance), and channel estimation. For each category, we provide detailed reviews of current studies. In particular, our major contributions are summarized as follows:
\begin{enumerate}
\item We examine the noise that may affect the performance of an estimation scheme and the metrics that are usually used to assess the performance of an estimation scheme.
\item We provide a detailed review on different distance estimation schemes. Then we compare the performance of different methods via numerical results by calculating the mean squared error (MSE).
\item We divide other parameters into environmental parameters, synchronization-related parameters, and SNR. Then we provide detailed reviews on each of them. Moreover, we review the pilot-based estimation scheme and the semi-blind CIR estimation scheme for channel estimation, and compare the performance of these two schemes. 
\item We identify and discuss promising future research directions for parameter estimation and channel estimation.		
\end{enumerate}

The rest of this paper is organized as follows. In Section \ref{mcvd}, we introduce a molecular communication via diffusion (MCvD) environment and present a summary of the noise that may impact on the estimation performance. In Section \ref{pmpe}, we present metrics for evaluating the performance of an estimation scheme. In Section \ref{de}, we review the distance estimation schemes in current studies and compare the performance of different estimation schemes. In Section \ref{pe}, we review the estimation of other parameters. In Section \ref{ce}, we review channel estimation. In Section \ref{frd}, we present future research directions for parameter estimation and channel estimation. Conclusions are drawn in Section \ref{c}.

\section{Molecular Communication via Diffusion}\label{mcvd}

\begin{figure}[!t]
    \begin{center}
	\includegraphics[height=1.5in,width=0.8\columnwidth]{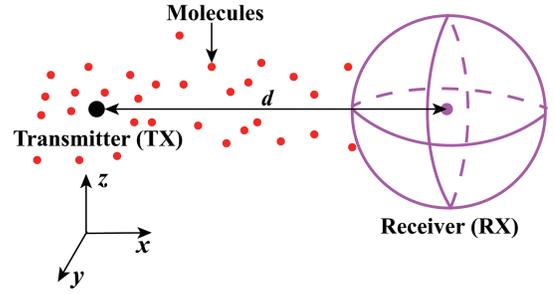}
    \caption{Illustration of the MCvD environment where one point $\mathrm{TX}$ communicates with one spherical $\mathrm{RX}$ in a three-dimensional environment.}\label{sys3d}\vspace{-0.5em}
    \end{center}\vspace{-4mm}
\end{figure}

As most studies considered estimation schemes in an MCvD system\footnote{We also review studies that perform estimation in a non-MCvD environment in Section \ref{Ms}.}, we provide an explanation of this system as shown in Fig. \ref{sys3d}, where one point $\mathrm{TX}$ communicates with one spherical $\mathrm{RX}$ with radius $r_{\ss\mathrm{R}}$ in an unbounded three-dimensional (3D) environment. The $\mathrm{RX}$ center is at a distance $d$ away from the $\mathrm{TX}$. We also consider the $\mathrm{TX}$ as a point source which releases molecules into the environment. 

\subsection{Propagation Channel Modeling}

We assume that the propagation channel outside the $\mathrm{TX}$ and the $\mathrm{RX}$ is filled with a fluid medium. Once molecules are released, they diffuse randomly in the propagation environment. The movement speed of molecules is determined by the diffusion coefficient, denoted by $D$, that is affected by the temperature of the fluid medium, the dynamic viscosity of the fluid, and the Stoke's radius of molecules. In this paper, we assume that the fluid medium has uniform temperature and viscosity such that $D$ can be modeled as a constant value. It is noted that a more complex MC environment can incorporate flow with a constant velocity $v$ and degradation of molecules, i.e., molecules of type $A$ can degrade into some other molecular species $\hat{A}$ with a constant degradation rate $k$. $\hat{A}$ cannot be recognized by the $\mathrm{RX}$.

\subsection{Receiver Modeling}

Current studies on estimation schemes focused on three types of $\mathrm{RXs}$, i.e., transparent $\mathrm{RX}$, fully-absorbing $\mathrm{RX}$, and reactive $\mathrm{RX}$. In this paper, we denote $h(t)$ as the CIR of the end-to-end channel.

\subsubsection{Transparent RX Modeling}\label{ss}

The transparent $\mathrm{RX}$ does not impede the diffusion of molecules, nor interact with molecules. We count the number of free molecules that are within the $\mathrm{RX}$ volume as the received signal. The CIR of the transparent $\mathrm{RX}$ is given by \cite[eq. (4)]{noel2014improving}
\begin{align}\label{htt}
h(t)=\frac{V_{\ss\mathrm{RX}}}{\left(4\pi Dt\right)^\frac{3}{2}}\exp\left(-\frac{d^2}{4Dt}\right),
\end{align}
where $V_{\ss\mathrm{RX}}$ is the volume of the $\mathrm{RX}$ and $V_{\ss\mathrm{RX}}=\frac{4}{3}\pi r_{\ss\mathrm{R}}^3$ for the spherical $\mathrm{RX}$. It is noted that $h(t)$ incorporating the flow and degradation of molecules is given by \cite[eq. (13)]{noel2014optimal}. It is also noted that \eqref{htt} is accurate when the $\mathrm{RX}$ is sufficiently far away from the $\mathrm{TX}$, i.e., $d$ is very large relative to $r_{\ss\mathrm{R}}$. Thus, it is reasonable to assume that the concentration of molecules at every point within the $\mathrm{RX}$ equals the concentration at the central point of the $\mathrm{RX}$. If the $\mathrm{RX}$ is close to the $\mathrm{TX}$, the uniform assumption of concentration does not hold. In this case, $h(t)$ is given by \cite[eq. (27)]{noel2013using}
\begin{align}\label{uht}
	h(t)&=\frac{1}{2}\left(\mathrm{erf}\left(\frac{r_{\ss\mathrm{R}}-d}{\sqrt{4Dt}}\right)+\mathrm{erf}\left(\frac{r_{\ss\mathrm{R}}+d}{\sqrt{4Dt}}\right)\right)+\frac{\sqrt{Dt}}{r_{\ss\mathrm{R}}\sqrt{\pi}}\notag\\&\times\left(\exp\left(-\frac{(r_{\ss\mathrm{R}}-d)^2}{4Dt}\right)+\exp\left(-\frac{(r_{\ss\mathrm{R}}+d)^2}{4Dt}\right)\right),
\end{align}
where $\mathrm{erf}(\cdot)$ denotes the error function. We note that \eqref{htt} is more widely applied in existing studies than \eqref{uht}, due to its simplicity, and \eqref{htt} provides an accurate approximation for \eqref{uht} if $r_{\ss\mathrm{R}}<0.15d$ \cite{noel2013using}.

\begin{figure}[!t]
    \begin{center}
    \includegraphics[height=2.2in,width=0.8\columnwidth]{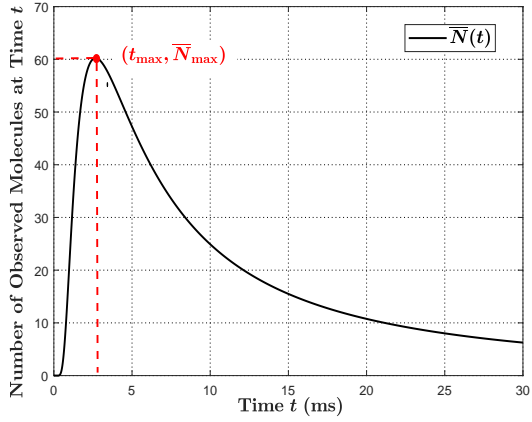}
    \caption{The number of observed molecules within the transparent $\mathrm{RX}$ at time $t$ versus time $t$, where $N_\mathrm{tx}=10^5$,  $r_{\ss\mathrm{R}}=0.5\;\mu\mathrm{m}$, $d=4\;\mu\mathrm{m}$, and $D=1000\;\mu\mathrm{m}^2/\mathrm{s}$ \cite{noel2014bounds}.}\label{peak}\vspace{-0.5em}
	\end{center}\vspace{-4mm}
\end{figure}

We now denote $\overline{N}(t)$ as the expected number of molecules observed within the $\mathrm{RX}$ volume at time $t$. If an impulse of $N_\mathrm{tx}$ molecules is released from the $\mathrm{TX}$ at time $t_0=0$, $\overline{N}(t)$ is given by $\overline{N}(t)=N_\mathrm{tx}h(t)$. In Fig. \ref{peak}, we plot $\overline{N}(t)$ versus time $t$ by adopting \eqref{htt}. From this figure, we observe a peak number of molecules observed within the $\mathrm{RX}$. We denote the time for reaching the peak number of molecules observed as $t_\mathrm{max}$. By taking the derivative of \eqref{htt} with respective to $t$, $t_\mathrm{max}$ is calculated as \cite[eq. (6)]{noel2014bounds}
\begin{align}\label{tmax}
t_\mathrm{max}=\frac{d^2}{6D}.
\end{align}
By substituting \eqref{tmax} into \eqref{htt}, we obtain the peak CIR, denoted by $h_\mathrm{max}$, as
\begin{align}\label{htm}
h_\mathrm{max}
=\left(d\sqrt{\frac{2\pi e}{3}}\right)^{-3}V_\mathrm{RX}.
\end{align}
We denote $\overline{N}_\mathrm{max}$ as the expected peak number of molecules observed within the RX. Then $\overline{N}_\mathrm{max}$ is given by $\overline{N}_\mathrm{max}=N_\mathrm{tx}h_\mathrm{max}$.

\subsubsection{Fully-Absorbing $\mathrm{RX}$ Modeling}	

In biological systems, many practical $\mathrm{RX}$ surfaces may interact with the molecules of interest, e.g., by providing binding sites for absorption or other reactions \cite{cuatrecasas1974membrane}. Hence, the transparent $\mathrm{RX}$ model is over-simplified. One practical $\mathrm{RX}$ model is the fully-absorbing $\mathrm{RX}$. In this model, the $\mathrm{RX}$ absorbs molecules as soon as they hit the surface. The fully-absorbing $\mathrm{RX}$ counts the total number of molecules absorbed as the received signal, where the CIR is given by \cite[eq. (23)]{yilmaz2014three}
\begin{align}\label{fht}
	h(t)=\frac{r_{\ss\mathrm{R}}}{d}\mathrm{erfc}\left(\frac{d-r_{\ss\mathrm{R}}}{\sqrt{4Dt}}\right).
\end{align}

\subsubsection{Reactive $\mathrm{RX}$ Modeling}
Different from the fully-absorbing $\mathrm{RX}$, molecules that reach the $\mathrm{RX}$ may participate in a reversible bimolecular second-order reaction with receptors over the $\mathrm{RX}$ membrane. This type of $\mathrm{RX}$ is named as the reactive $\mathrm{RX}$ whose received signal is the number of activated receptors. The CIR of the reactive $\mathrm{RX}$ is given by \cite[eq. (29)]{ahmadzadeh2016comprehensive} and is omitted here due to its complex format.

\subsection{Noise Modeling}

We next review some factors that affect the performance of estimation and treat these factors as noise during the estimation process.

\subsubsection{Statistical Distribution of Received Signal}

Due to the random diffusion (RD) of molecules, the number of molecules observed at the $\mathrm{RX}$ is a random variable (RV) \cite{pierobon2011diffusion}. This randomness influences the performance of estimation. Due to the independent diffusion of molecules, any given molecule released by the $\mathrm{TX}$ is observed by the $\mathrm{RX}$ with a probability of $h(t)$. A binary state model applies and the number of molecules observed at time $t$, denoted by $N_\mathrm{ob}(t)$, follows a binomial distribution with $N_\mathrm{tx}$ trails and success probability $h(t)$. This is mathematically expressed as
\begin{align}
N_\mathrm{ob}(t)\sim\mathcal{B}(N_\mathrm{tx},h(t)),
\end{align}
where $\mathcal{B}(N,p)$ represents a binomial distribution. Unfortunately, the binomial distribution is cumbersome to work with in MC systems. Therefore, current studies usually approximate binomial distribution as two distributions for the sake of mathematical tractability, described as follows:
\begin{itemize}
\item[(a)] \textit{Poisson distribution}: When the number of trials $N_\mathrm{tx}$ is large and the success probability $h(t)$ is small, $N_\mathrm{ob}(t)$ can be approximated as a Poisson RV, given by
	\begin{align}\label{po}
	N_\mathrm{ob}(t)\sim\mathcal{P}\left(N_\mathrm{tx}h(t)\right),
	\end{align}
    where $\mathcal{P}(\varphi)$ represents the Poisson distribution with the mean of $\varphi$. Based on \eqref{po}, the probability mass function (PMF) of the Poisson RV $N_\mathrm{ob}(t)$ is written as
    \begin{align}
    \mathrm{Pr}\left(N_\mathrm{ob}(t)=\xi\right)=
    \frac{\left(N_\mathrm{tx}h(t)\right)^{\xi}
    \exp\left(-N_\mathrm{tx}h(t)\right)}{\xi!},
	\end{align}
	where $\mathrm{Pr}(\cdot)$ stands for the probability.
\item[(b)] \textit{Gaussian distribution}: If the expected number of molecules observed, i.e., $\overline{N}(t)$, is sufficiently large, we can apply the central limit theorem and approximate $N_\mathrm{ob}(t)$ as a Gaussian RV, given by
	\begin{align}
	N_\mathrm{ob}(t)\sim\mathcal{N}\left(N_\mathrm{tx}h(t), N_\mathrm{tx}h(t)(1-h(t))\right).
	\end{align}
    The probability density function (PDF) of $N_\mathrm{ob}(t)$ is given by
    \begin{align}
    \mathrm{Pr}\left(N_\mathrm{ob}(t)=\xi\right)=\frac{\exp\left(-\frac{(\xi-N_\mathrm{tx}h(t))^2}
    {2N_\mathrm{tx}h(t)(1-h(t))}\right)}{\sqrt{2\pi N_\mathrm{tx}h(t)(1-h(t))}}.
    \end{align}
\end{itemize}

\subsubsection{External Additive Noise}

In MC systems, the intended $\mathrm{TX}$ may not be the only source of molecules. Other sources, referred to as external sources, may also release the same type of molecules that influence the observation at the $\mathrm{RX}$ and affects the estimation performance. We detail some examples of external sources as follows:
\begin{itemize}
\item[(a)] \textit{Multiuser interference:} Noisy molecules are emitted by $\mathrm{TXs}$ in other MC systems.
\item[(b)] \textit{Unintended leakage:} Molecules can be leaked from membrane-bound containers, e.g., vesicles, within transceivers. A rupture can result in a steady or sudden release of molecules \cite{ladokhin1995leakage}.
\item[(c)] \textit{Output from unrelated biochemical processes}: Biocompatibility of the MC system may require the selection of naturally-occurring molecules. Therefore, other processes that produce the same type of molecules are noisy sources for the considered MC system. For example, calcium is commonly used as a messenger molecule within cellular systems \cite[Ch. 16]{alberts2015essential}. If calcium is applied as signaling molecules of the MC system in the biological environment, the naturally-occurring calcium would impact the MC system.
\item[(d)] \textit{Unintended reception of other molecules:} Molecules that are highly similar to intended molecules may be recognized by the $\mathrm{RX}$. For example, the receptors at the $\mathrm{RX}$ may bind to other molecules that have a very similar shape and size to intended molecules \cite[Ch. 4]{alberts2015essential}.
\end{itemize}

We denote $N_\mathrm{sig}(t)$ as the intended observed molecules and $n(t)$ as the observed noise molecules. Since intended molecules and noise molecules are indistinguishable, the total number of molecules observed at the $\mathrm{RX}$ at time $t$ is given by
\begin{align}
N_\mathrm{ob}(t)=N_\mathrm{sig}(t)+n(t).
\end{align}
The analysis of the statistical distribution of $n(t)$ is built upon following assumptions:
\begin{itemize}
\item[A1)] We denote $\overline{n}$ as the expected number of noise molecules observed within the $\mathrm{RX}$. We assume that $\overline{n}$ is constant over the entire observation time.
\item[A2)] The observation of one noise molecule at the $\mathrm{RX}$ is independent of observations of other noise molecules.
\item[A3)] The uniform concentration assumption holds for noise molecules at the $\mathrm{RX}$.
\end{itemize}
Based on A1)--A3), we model the number of observed noise molecules as a Poisson RV, due to the law of rare events (LRE) \cite{falk2010laws}, i.e., $n(t)\sim\mathcal{P}(\overline{n})$.

\subsubsection{ISI \& ILI}

ISI exists when the $\mathrm{TX}$ transmits multiple symbols to the $\mathrm{RX}$. Due to the RD of molecules, the molecules from previously sent symbols may arrive at the $\mathrm{RX}$ in the current symbol interval, which influences the estimation in the current symbol interval. ILI exists for the multiple-input multiple-output (MIMO) MC system, where one $\mathrm{TX}-\mathrm{RX}$ channel is influenced by molecules released from other channels.

\section{Performance Metrics for Estimation Schemes}\label{pmpe}

In this section, we review some metrics that are usually applied to evaluate the performance of an estimation scheme.

\subsection{MSE}

The MSE is usually applied to assess the quality of an estimation scheme. We denote $\theta$ as the unknown parameter and $\hat{\theta}$ as the estimated value of $\theta$. The MSE of an estimation scheme is defined as \cite{pishro2016introduction}
\begin{align}
\mathrm{MSE}(\hat{\theta})=\E\left[(\hat{\theta}-\theta)^2\right],
\end{align}
where $\E\left[\cdot\right]$ represents the expectation. The MSE can be written as the sum of the variance and squared bias of the estimation scheme, which is \cite{wackerly2014mathematical}
\begin{align}
\mathrm{MSE}(\hat{\theta})=\mathrm{Var}\left(\hat{\theta}\right)
+\mathrm{Bias}\left(\hat{\theta},\theta\right)^2,
\end{align}
where the variance is $\mathrm{Var}(\hat{\theta})=\E[(\hat{\theta}-\E[\hat{\theta}])^2]$ and the squared bias is $\mathrm{Bias}(\hat{\theta},\theta)^2=(\E[\hat{\theta}]-\theta)^2$. For any unbiased estimation scheme, $\E[\hat{\theta}]=\theta$. Thus, the MSE equals the variance.

\subsection{Cramer-Rao Lower Bound (CRLB)}

The CRLB is a lower bound on the variance of any unbiased estimation scheme \cite[Ch. 3]{kay1993fundamentals}. An estimation scheme is called the minimum-variance unbiased (MVU) estimator if its MSE attains the CRLB. Therefore, the CRLB provides insights into the comparison of estimation schemes and the prediction of the performance of the MVU estimator.

We consider an $M$-point data set $\mathbf{x}=\left[x_1, x_2,\cdots,x_M\right]$ that depends on an unknown parameter vector $\boldsymbol{\theta}$, where $\boldsymbol{\theta}$ contains $L$ unknown parameters as $\boldsymbol{\theta}=[\theta_1, \theta_2,\cdots,\theta_L]$. Hence, the date set $\mathbf{x}$ is applied to determine $\boldsymbol{\theta}$. To mathematically model the data set, we define a PDF as $p(\mathbf{x};\boldsymbol{\theta})$ that is parameterized by the unknown parameter vector $\boldsymbol{\theta}$. For the CRLB to exist, the regularity condition must be satisfied, which is given by \cite[Ch. 3]{kay1993fundamentals}
\begin{align}\label{regu}
	\E\left[\frac{\partial\ln p(\mathbf{x};\boldsymbol{\theta})}{\partial \boldsymbol{\theta}}\right]=\mathbf{0},~\mathrm{for}~\mathrm{all}~\boldsymbol{\theta},
\end{align}
where the expectation is taken with respect to $p(\mathbf{x};\boldsymbol{\theta})$. We denote $\hat{\boldsymbol{\theta}}$ as the estimated parameter vector of $\boldsymbol{\theta}$. The covariance matrix of any unbiased estimation scheme $\hat{\boldsymbol{\theta}}$, denoted by $\mathbf{C}_{\hat{\boldsymbol{\theta}}}$, satisfies\cite[eq. (3.24)]{kay1993fundamentals}
\begin{align}		\mathbf{C}_{\hat{\boldsymbol{\theta}}}-\mathbf{I}^{-1}(\boldsymbol{\theta})\geq\mathbf{0},
\end{align}
where $\mathbf{I}(\boldsymbol{\theta})$ is an $L\times L$ Fisher information matrix, given by
\begin{align}\label{fim}
	\left[\mathbf{I}(\boldsymbol{\theta})\right]_{ij}=-\E\left[\frac{\partial^2\ln p(\mathbf{x};\boldsymbol{\theta})}{\partial\theta_i\partial\theta_j}\right],
\end{align}
with $i=1,2,\cdots,L$ and $j=1,2,\cdots,L$. The derivatives are evaluated at the true value of $\boldsymbol{\theta}$. The CRLB on $\theta_l$ is found as the $[l, l]$ element of the inverse of $\mathbf{I}(\boldsymbol{\theta})$, which is
\begin{align}
	\mathrm{Var}(\hat{\theta}_l)\geq\left[\mathbf{I}^{-1}(\boldsymbol{\theta})\right]_{ll}.
\end{align}

When only a single papermeter $\theta$ is unknown, \eqref{regu} is simplified as 
\begin{align}\label{exp}
	\E\left[\frac{\partial\ln p(\mathbf{x};\theta)}{\partial\theta}\right]=0,~\mathrm{for}~ \mathrm{all}~\theta,
\end{align}
where $p(\mathbf{x};\theta)$ is the PDF parameterized by the unknown parameter $\theta$. We simplify \eqref{fim} as \cite[eq. (3.6)]{kay1993fundamentals}
\begin{align}\label{fi}
	I(\theta)=-\E\left[\frac{\partial^2\ln p(\mathbf{x};\theta)}{\partial\theta^2}\right],
\end{align}
where $I(\theta)$ is the Fisher information. The CRLB for the unbiased estimation of $\theta$ is $\mathrm{Var}(\hat{\theta})\geq I^{-1}(\theta)$.

\subsection{Hammersley-Chapman-Robbins Lower Bound (HCRLB)}

The HCRLB is a lower bound on the variance of any unbiased estimation scheme for a function parameterized by the unknown parameter \cite{lehmann2006theory}. Compared to the CRLB, the HCRLB is tighter and does not need to satisfy the regularity condition, while the computation is more complex. We denote $g(\theta)$ as a function of the unknown parameter $\theta$ and $\hat{g}(\theta)$ as an estimated function of $g(\theta)$. The HCRLB on the variance of $\hat{g}(\theta)$ is given by
\begin{align}\label{hcrlb}
\mathrm{Var}\left(\hat{g}(\theta)\right)\geq\underset{\Delta}
{\mathrm{sup}}\frac{\left[g(\theta+\Delta)-g(\theta)\right]^2}
{\E\left[\frac{p(\mathbf{x};\theta+\Delta)}{p(\mathbf{x};\theta)}-1\right]^2},
\end{align}
where $\mathrm{sup}$ stands for supremum. If we set $g(\theta)=\theta$ and substitute it into \eqref{hcrlb}, we obtain the HCRLB on $\mathrm{Var(\hat{\theta})}$. In \eqref{hcrlb}, the HCRLB converges to the CRLB when $\Delta\rightarrow 0$. The HCRLB can be applied to a wider range of problems. For example, if $p(\mathbf{x};\theta)$ is non-differentiable, the Fisher information is not defined. Hence, the CRLB does not exist. However, the HCRLB may exist under this condition.

\section{Distance Estimation}\label{de}

In this section, we review the current studies on the estimation of the distance between the $\mathrm{TX}$ and the $\mathrm{RX}$ in MC systems. We classify the current studies about distance estimation into two-way estimation and one-way estimation in Section \ref{tw} and Section \ref{ow}, respectively. We summarize different distance estimation schemes in Table \ref{table1}. It is noted that some studies focused on the one-dimensional (1D) environment while other studies focused on the 3D environment. To facilitate the comparison, we consider all estimation schemes in the 3D environment via following the estimation process and replacing the CIR in 1D with $h(t)$ given in \eqref{htt}, \eqref{uht}, or \eqref{fht} for different types of $\mathrm{RXs}$.

\begin{table*}[!ht]
\newcommand{\tabincell}[2]{\begin{tabular}{@{}#1@{}}#2\end{tabular}}
\centering
\caption{Distance Estimation Schemes for MC Systems}\label{table1}
	\scalebox{0.96}{\begin{tabular}{|c|c|c|c|c|c|c|c|c|}\hline				\textbf{Name}&\textbf{References}&\textbf{Performance}&\textbf{\tabincell{c}{Computational\\ Complexity}}&\textbf{Synchronization}&\textbf{Noise}&\textbf{\tabincell{c}{TX\\ Waveform}}&\textbf{RX Type}&\textbf{Environment}\\
\hline
RTT&\cite{moore2012measuring,moore2012comparing}&Moderate&Low&Not Required&RD&Impulse&Transparent&1D \\
\hline
SA&\cite{moore2012measuring}&Low&Low&Not Required&RD&Impulse&Transparent&1D\\
\hline
\multirow{4}{*}{\tabincell{c}{Peak-Based\\(One Type of\\ Molecule)}}&\cite{huang2013distance,liu2019localization}&Moderate&Low&Required&RD&Impulse&Transparent&1D, 2D\\
\cline{2-9}				&\cite{wang2015distance}&Moderate&Low&Required&RD&Rectangular&Transparent&3D\\
\cline{2-9}				&\cite{turan2018transmitter}&Moderate&Low&Required&RD&Impulse&\tabincell{c}{Transparent,\\Ring-Shaped}&\tabincell{c}{Cylindrical,\\Poiseuille Flow}\\
\hline
{\tabincell{c}{Peak-Based\\(Two Types of\\ Molecules)}}&\cite{luo2018effective,sun2017efficient}&Low&Low&Not Required&RD&Impulse&Transparent&1D\\
\hline				\multirow{4}{*}{ML}&\cite{noel2014bounds,kumar2020nanomachine}&High&High&Required&RD&Impulse&Transparent&\tabincell{c}{3D, Flow,\\Molecules\\ Degradation}\\
\cline{2-9}
&\cite{lin2019high}&High&High&Required&ISI&Impulse&Transparent&3D\\
\cline{2-9}				&\cite{huang2020initial}&High&High&Required&RD&\tabincell{c}{Impulse,\\Diffusive}&\tabincell{c}{Transparent,\\Diffusive}&3D\\
\hline
\tabincell{c}{Fraction of\\Absorbed Molecules}&\cite{wang2015algorithmic}&N/A&High&\tabincell{c}{Not Required\\\&Required}&ISI, RD&Impulse&Fully-Absorbing&3D\\
\hline
Data  Fitting&\cite{miao2019cooperative}&N/A&High&Required&RD&Impulse&\tabincell{c}{Multiple,\\Fully-Absorbing}&3D\\
\hline
Macro-scale&\cite{gulec2020distance}&N/A&High&Required&N/A&Sprayer&MQ-3 Sensor&Tabletop\\
\hline
\end{tabular}}
\end{table*}

\subsection{Two-Way Estimation}\label{tw}

Two-way estimation schemes were proposed in \cite{moore2012measuring} to estimate the distance between two transceivers that are labeled as $T$ and $R$, respectively. Specifically, $T$ first releases an impulse of molecules of type $A$ at time $t_0$. The diffusion coefficient of molecules $A$ is $D_{A}$. When $R$ detects the molecules $A$ at time $t_1$, it immediately transmits a feedback signal of an impulse of molecules $B$ whose diffusion coefficient is $D_{B}$. At time $t_2$, $T$ detects the molecules $B$. $T$ and $R$ are regarded as transparent $\mathrm{RXs}$ when they detect molecules $B$ and $A$, respectively.

\subsubsection{Round Trip Time (RTT) Protocols}

In RTT protocols, $T$ measures the RTT that is the sum of time required for the transmission from $T$ to $R$ and for the transmission from $R$ to $T$. The first estimation scheme is named as \textit{the RTT protocol from peak concentration (RTT-P)}. In RTT-P, $T$ transmits at time $t_0$ and $R$ detects the peak concentration of molecules $A$ from $T$ at time $t_1$. According to \eqref{tmax}, we can obtain the relationship between $t_1-t_0$ and $d$. $T$ detects the peak concentration of the feedback signal with type $B$ molecules at time $t_2$. Similarly, we can obtain the relation between $t_2-t_1$ and $d$ based on \eqref{tmax}. Accordingly, the distance $d$ is estimated as a function of the RTT $t_2-t_0$ as
\begin{align}\label{1}
\hat{d}=\sqrt{\frac{6 D_{A}D_{\ss B}}{D_{A}+D_{B}}(t_2-t_0)},
\end{align}
where $\hat{d}$ is the estimated value of $d$.

The second estimation scheme was named as \textit{the RTT protocol from threshold concentration (RTT-T)}. Different from RTT-P, RTT-T defines threshold concentrations $H_{A}$ and $H_B$ for $R$ and $T$ to detect the number of molecules observed, respectively. It is assumed that $T$ transmits $N_\mathrm{tx}^A$ number of molecules $A$ and $R$ transmits $N_\mathrm{tx}^B$ number of molecules $B$. $R$ records $t_1$ when the number of molecules observed reach the threshold concentration $H_A$, i.e., $\overline{N}(t_1-t_0)|_{D=D_A, N_\mathrm{tx}=N_\mathrm{tx}^A}=H_A$ by assuming $\overline{N}(t)=N_\mathrm{ob}(t)$. Similarly, $T$ records $t_2$ when the number of molecules observed reach the threshold concentration $H_B$, i.e., $\overline{N}(t_2-t_1)|_{D=D_B, N_\mathrm{tx}=N_\mathrm{tx}^B}=H_B$. If $D_A=D_B$, $N_\mathrm{tx}^A=N_\mathrm{tx}^B$, and $H_A=H_B$, $d$ is estimated as a function of the RTT $t_2-t_0$ as
\begin{align}\label{2}
\hat{d}=\sqrt{D_A(t_2-t_0)\ln\left(\frac{V_\mathrm{RX}^2
(N_\mathrm{tx}^A)^2}{8\pi^3D_A^3H_A^2(t_2-t_0)^3}\right)}.
\end{align}

\subsubsection{Signal Attenuation Protocol from Peak Concentration (SA-P)}

In SA-P, $T$ transmits type $A$ molecules to $R$, and $R$ measures the peak concentration, denoted by $N_\mathrm{ob,m}^A$. According to \eqref{htm}, we can obtain the relationship between $N_\mathrm{ob,m}^A$ and $d$ by replacing $\overline{N}_\mathrm{max}$ with $N_\mathrm{ob,m}^A$. Similarly, $R$ transmits type $B$ molecules and $T$ measures the peak concentration, denoted by $N_\mathrm{ob,m}^B$. By assuming $N_\mathrm{tx}^B=N_\mathrm{ob,m}^A$, $\hat{d}$ is obtained as
\begin{align}\label{3}
\hat{d}=\sqrt{\frac{3}{2\pi e}}\left(\frac{N_\mathrm{tx}^AV_\mathrm{RX}^2}{N_\mathrm{ob,m}^B}\right)^\frac{1}{6}.
\end{align}	

The estimation performance of SA has been shown in \cite{noel2014bounds}.

\subsubsection{Merits and Drawbacks}\label{MD}

In this subsection, we summarize the merits and drawbacks of the two-way estimation. One merit of this estimation is that the synchronization between two transceivers is not required. For both RTT and SA protocols, only the time period of two-way transmission, the number of emitted molecules, and peak observed concentration at $T$ are required. Despite this merit, there are several drawbacks. The first drawback is that this estimation is time-consuming since it requires two-way transmission. The second one is that this estimation requires $R$ to immediately send the feedback signal when it detects the type of molecules $A$, which is challenge for a nanomachine. The third drawback is that the instantaneous observation at the $\mathrm{RX}$ is used to approximate its expectation, which affects the estimation performance.

\subsection{One-Way Estimation}\label{ow}

Due to the aforementioned drawbacks of the two-way estimation, most current studies have focused on the acquisition of the distance information from the received signal only, referred to as one-way estimation.

\subsubsection{Peak-Based Estimation}

In this subsection, we review the studies that perform the estimation via measuring the peak received signal or the time for reaching the peak received signal.

First, \cite{huang2013distance} performed the estimation based on the peak number of observed molecules, denoted by $N_\mathrm{ob,m}$, at the transparent $\mathrm{RX}$. According to \eqref{htm}, the distance $d$ is estimated as
\begin{align}
\hat{d}=\sqrt{\frac{3}{2\pi e}}\left(\frac{V_\mathrm{RX}N_\mathrm{tx}}{N_\mathrm{ob,m}}\right)^\frac{1}{3}.
\end{align}
The estimation performance of this method is affected by the RD of molecules, since the instantaneous observation is used to approximate its expectation.

Second, \cite{wang2015distance} considered, for the first time, the release of molecules from the $\mathrm{TX}$ as a rectangular pulse, where the CIR at the transparent $\mathrm{RX}$ is denoted by $h_\mathrm{rec}(t)$. By taking the derivative of $h_\mathrm{rec}(t)$ with respect to $t$, $d$ can be estimated via measuring the time for reaching the peak concentration at the $\mathrm{RX}$ as
\begin{align}\label{tm}
\hat{d}=\sqrt{\frac{6Dt_\mathrm{max}\left(t_\mathrm{max}-T_\mathrm{e}\right)}{T_\mathrm{e}}
\ln\left(\frac{t_\mathrm{max}}
{t_\mathrm{max}-T_\mathrm{e}}\right)},
\end{align}
where $T_\mathrm{e}$ is the emission duration.

As $t_\mathrm{max}$ is involved in the estimation of \cite{wang2015distance}, a perfect synchronization between the $\mathrm{TX}$ and the $\mathrm{RX}$ is required. In \cite{luo2018effective,sun2017efficient}, the authors proposed a low-complexity scheme that does not require the synchronization via adopting two types of molecules. In this scheme, the $\mathrm{TX}$ releases type $A$ molecules at time $t_0$ and the $\mathrm{RX}$ records the time of peak concentration of molecules $A$, denoted by $t_\mathrm{max}^A$. Similarly, the $\mathrm{TX}$ transmits type $B$ molecules at $t_1$, and the $\mathrm{RX}$ records the time of peak concentration of molecules $B$, denoted by $t_\mathrm{max}^B$. According to \eqref{tmax}, the relation between $t_\mathrm{max}^A$ and $d$, and $t_\mathrm{max}^B$ and $d$ are obtained. Thus, $d$ is estimated as
\begin{align}
\hat{d}=\sqrt{\frac{6D_AD_B\left(\Delta t_{\ss\mathrm{RX}}-\Delta t_{\ss\mathrm{TX}}\right)}{D_A-D_B}},
\end{align}
where $\Delta t_{\ss\mathrm{RX}}=t_\mathrm{max}^B-t_\mathrm{max}^A$ and $\Delta t_{\ss\mathrm{TX}}=t_1-t_0$. We note that $\Delta t_{\ss\mathrm{RX}}$ and $\Delta t_{\ss\mathrm{TX}}$ are based on the time measurements at the $\mathrm{RX}$ and the $\mathrm{TX}$, respectively. Therefore, synchronization is not required.

\begin{figure}[!t]
    \begin{center}
	\includegraphics[height=1.5in,width=1\columnwidth]{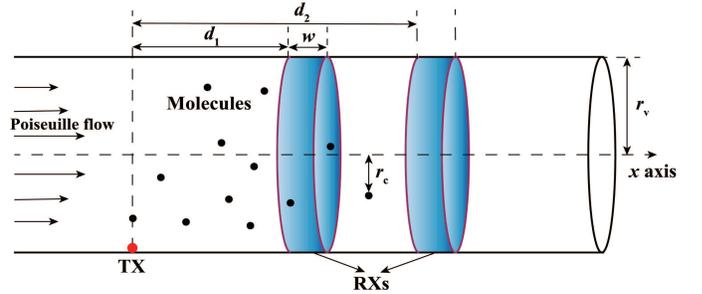}
    \caption{Estimation in a cylindrical diffusive MC environment by using ring-shaped $\mathrm{RXs}$; reproduced based on \cite{turan2018transmitter}.}
    \label{vessel}\vspace{-0.5em}
	\end{center}\vspace{-4mm}
\end{figure}

In \cite{moore2012measuring,huang2013distance,wang2015distance,luo2018effective,sun2017efficient}, estimation is performed in an unbounded environment. Different from these studies, \cite{turan2018transmitter} considered a cylindrical MC environment whose surface is reflecting as shown in Fig. \ref{vessel}. Within the cylindrical environment, a point $\mathrm{TX}$ releases molecules and two ring-shaped $\mathrm{RXs}$ with a certain width $w$ are located on the cylinder perimeter with a radius $r_\mathrm{v}$ that is the same as the cylinder's radius. After the molecules are released, they diffuse randomly with a constant diffusion coefficient $D$ and subject to the Poiseuille flow \cite{bruus2008theoretical}. As the cylindrical MC environment is a good approximation of the blood vessel environment, it has been widely investigated in existing studies, e.g., \cite{turan2018transmitter,wicke2018modeling,schafer2020transfer}. Due to the existence of the Poiseuille flow, advection and diffusion can both transport molecules. In \cite{turan2018transmitter}, Turan et al. considered a diffusion-domain movement and only focused on the movement in the $x$ axis, i.e., the channel is approximated as a 1D environment. The authors performed the estimation in two scenarios. The first scenario is that the emission time of molecules is known. By measuring the time of peak concentration, the distance can be estimated by a single $\mathrm{RX}$ as
\begin{align}\label{d1}
\hat{d}_1\approx\frac{-w+\sqrt{w^2-8\left(wv_mt_\mathrm{max}
-2(v_mt_\mathrm{max})^2-4D_\mathrm{e}t_\mathrm{max}\right)}}{4},
\end{align}
where $D_\mathrm{e}$ is the effective diffusion coefficient and $v_m$ is the average flow velocity.
The second scenario is that the emission time is unknown. Thus, $t_\mathrm{max}$ is unknown and \eqref{d1} contains two unknown parameters, i.e., $d_1$ and $t_\mathrm{max}$. In this scenario, $d_1$ is estimated by using two $\mathrm{RXs}$. Similar to \eqref{d1}, an equation containing $d_1$ and $t_\mathrm{max}$ can be obtained at the second $\mathrm{RX}$. Thus, $d_1$ can be mathematically derived by jointly solving these two equations.

\subsubsection{Maximum Likelihood (ML) Estimation}

The ML estimation is to find $\hat{\theta}$ that maximizes the joint observation likelihood, which usually achieves high accuracy, but requires high computational complexity and the perfect synchronization between the $\mathrm{TX}$ and the $\mathrm{RX}$. The authors in \cite{noel2014bounds,kumar2020nanomachine} considered $x_m$, i.e., the $m$th data of the data set $\mathbf{x}$, as the number of molecules observed at the transparent $\mathrm{RX}$ at time $t_m$ when molecules are released at time $t_0$. It is assumed that each observation is independent and follows a Poisson distribution. Thus, $p(\mathbf{x};\theta)$ is given by
\begin{align}\label{p_x_theta}
p(\mathbf{x};\theta)=\prod_{m=1}^{M}\frac{\left(N_\mathrm{tx}h(t_m)\right)^{x_m}}{x_m!}
\exp\left(-N_\mathrm{tx}h(t_m)\right).
\end{align}
Using \eqref{p_x_theta}, $d$ is estimated by taking the partial derivative of $p(\mathbf{x};\theta)$ with respect to $d$ and setting it equal to 0. Moreover, \cite{noel2014bounds} derived the CRLB on the variance of $\hat{d}$.

It is noted that \cite{noel2014bounds,kumar2020nanomachine} only considered the transmission of one symbol from the $\mathrm{TX}$ to the $\mathrm{RX}$. Different from that, \cite{lin2019high} considered multiple symbols transmitted from the $\mathrm{TX}$ to the transparent $\mathrm{RX}$, where ISI exists and may impact the estimation performance. Assuming that the $\mathrm{RX}$ makes one observation in each symbol interval, $p(\mathbf{x};\theta)$ is the joint PDF of multiple observations. Based on $p(\mathbf{x};\theta)$, $d$ is estimated.

Previous studies have focused on the distance estimation in a static MC system. Instead of that, \cite{huang2020initial} considered estimating the initial distance, i.e., the distance between the $\mathrm{TX}$ and the $\mathrm{RX}$ at the initial moment, in a diffusive mobile MC scenario, where both $\mathrm{TX}$ and $\mathrm{RX}$ diffuse with constant diffusion coefficients $D_\mathrm{TX}$ and $D_\mathrm{RX}$, respectively. A novel two-step scheme was proposed to estimate the initial distance, denoted by $d_0$. We denote $d(t)$ as the distance between the $\mathrm{TX}$ and the $\mathrm{RX}$ at time $t$, and the PDF of $d(t)$ is given by \cite[eq. (13)]{huang2020initial}, where $d_0$ is a parameter of the PDF of $d(t)$. Moreover, \cite{huang2020initial} proved that the ML estimation based on the joint PDF of multiple values of $d(t)$ can be simplified as performing the estimation based on the PDF of the first value of $d(t)$, i.e., $d(t_1)$. Therefore, the first step is to estimate the stochastic distance $d(t_1)$, where the estimation scheme is similar to \cite{noel2014bounds}. The second step is to estimate $d_0$ by finding $\hat{d}_0$ that maximizes the PDF of $d(t_1)$.
	
As the closed-form expression for $\hat{d}$ is difficult to derive in the ML estimation, \cite{lin2019high} and \cite{huang2020initial} used the Newton-Raphson method, which is a method to find successively better approximations to the roots of a real-valued function, in order to calculate $\hat{d}$. 
\subsubsection{Non-transparent RXs}

In this section, we review estimation schemes in an environment with non-transparent $\mathrm{RXs}$. Wang et al. in \cite{wang2015algorithmic} proposed an estimation scheme by applying a fully-absorbing $\mathrm{RX}$. The $\mathrm{RX}$ performs estimation by counting the number of molecules absorbed within a time interval. Specifically, an algorithm was proposed when the $\mathrm{TX}$ and the $\mathrm{RX}$ are unsynchronized. Moreover, the authors considered two optimization methods to improve the performance of estimation. The first method is using molecules with a large diffusion coefficient and the second method is increasing the number of emitted molecules.

Different from \cite{wang2015algorithmic}, the authors in \cite{miao2019cooperative} performed the estimation in an environment with multiple fully-absorbing $\mathrm{RXs}$. As one fully-absorbing $\mathrm{RX}$ would impact molecules absorbed by other fully-absorbing $\mathrm{RXs}$, an accurate derivation for the number of molecules absorbed at each $\mathrm{RX}$ is cumbersome. Miao et al. in \cite{miao2019cooperative} adopted a curve fitting method to obtain the expression for the number of absorbed molecules. Specifically, \cite{miao2019cooperative} used the nonlinear least squares method for curve fitting to obtain the distance, where the Levenberg-Marquardt (LM) method \cite{more1978levenberg} is adopted.

\subsubsection{Macro-scale MC Systems}\label{Ms}

\begin{figure}[!t]
	\begin{center}
	\includegraphics[height=2in,width=\columnwidth]{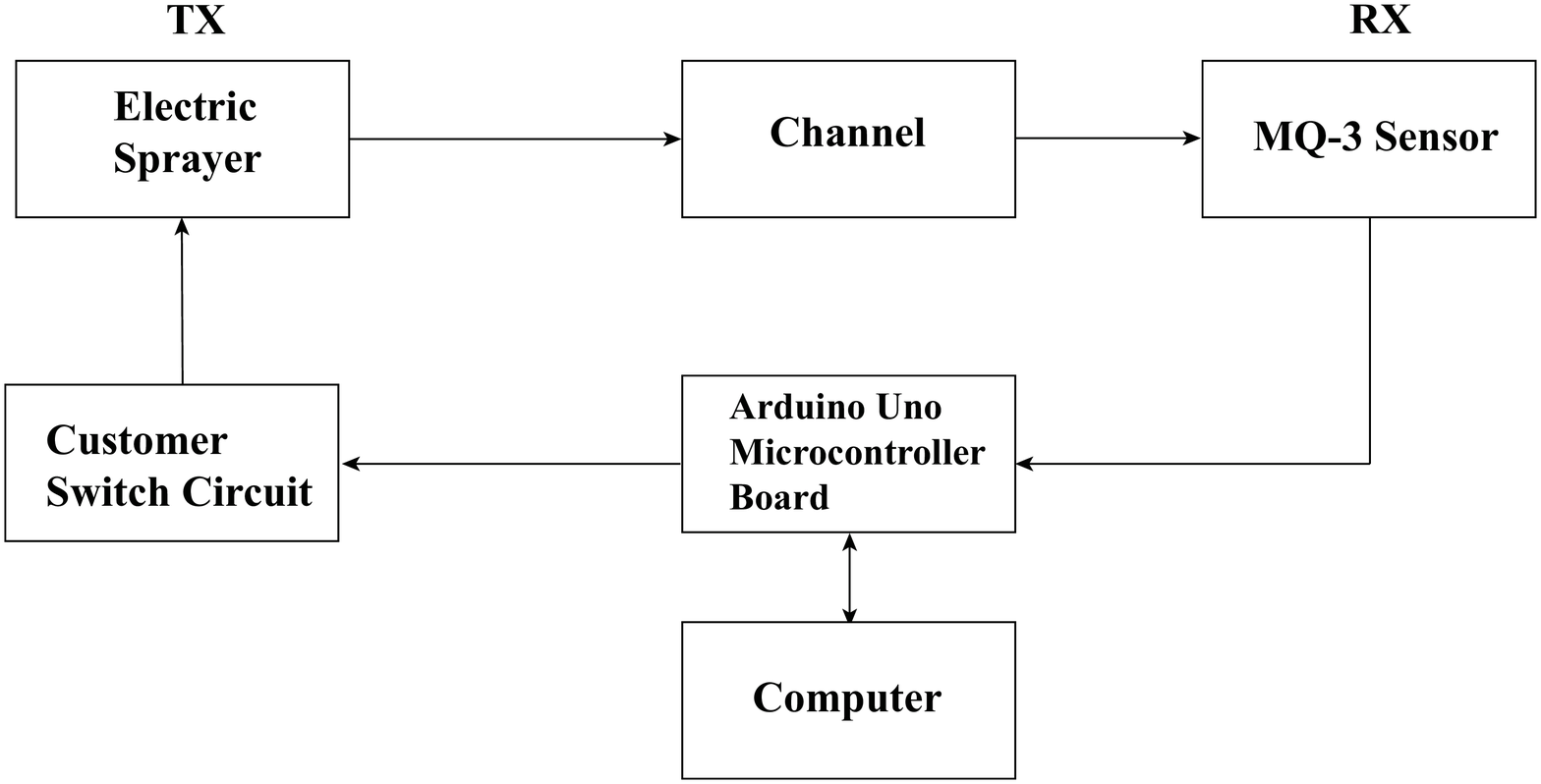}
    \caption{Block diagram of the experimental setup in \cite{gulec2020distance}.}\label{block}\vspace{-0.5em}
	\end{center}\vspace{-4mm}
\end{figure}

Previous studies focused on the distance estimation in micro-scale ($\mathrm{nm}$ to $\mu\mathrm{m}$) MC systems. With this focus, these studies have considered an ideal channel model where the transmitted molecules do not have an initial velocity, the molecules move according to Brownian motion, and the $\mathrm{TX}$ and the $\mathrm{RX}$ perfectly transmit and receive signals. In nature, MC also exists in the macro-scale ($\mathrm{cm}$ to $\mathrm{m}$) environment. For example, animals like bees, flies and ants use pheromone to send messengers over several meters. Against this background, \cite{gulec2020distance} investigated, for the first time, the distance estimation in the macro-scale environment. Instead of analyzing a theoretical model, \cite{gulec2020distance} established an experimental setup similar to the tabletop platform in \cite{farsad2013tabletop}. Fig. \ref{block} shows the block diagram of the experimental setup. The $\mathrm{TX}$ is an electric sprayer controlled by a micro-controller via a custom switch circuit to release ethyl alcohol molecules. The $\mathrm{RX}$ receives the molecular signal with an MQ-3 alcohol sensor. The $\mathrm{TX}$ and the $\mathrm{RX}$ are both controlled by an Arduino Uno micro-controller board which is connected to a computer. Five different practical methods were proposed for distance estimation at the $\mathrm{RX}$. The first two methods adopt the supervised machine learning, where multivariate linear regression and neural network regression are used. These methods use the extracted features, e.g., rise time on the rising edge of the measured signal, from the received molecular signals as inputs. The other three methods analyze the collected data, which are less complex but less accurate than machine learning methods. The first data analysis method employs the received power to estimate the distance. The second data analysis method employs the peak time of the received signal. The third method combines the power and peak time of the received signal to estimate the distance.

\subsection{Performance Comparison}

\begin{figure}[!t]
    \begin{center}
	\includegraphics[height=2in]{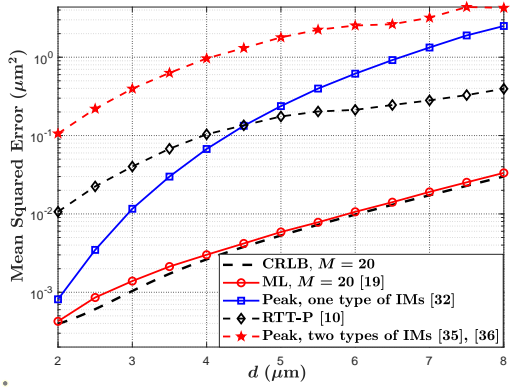}
    \caption{The MSE of different estimation schemes versus $d$.}\label{comparison}\vspace{-0.5em}
	\end{center}\vspace{-4mm}
\end{figure}

In this subsection, we present numerical results to compare the distance estimation performance of different estimation schemes via calculating their MSE. Particle-based simulation is used to simulate the random propagation of molecules \cite{andrews2004stochastic}. The simulation time step is $\Delta t_\mathrm{sim}=0.0001\;\mathrm{s}$ and all results are averaged over 10,000 realizations. Throughout this subsection, we set $N_\mathrm{tx}=N_\mathrm{tx}^A=N_\mathrm{tx}^B=10^5$, $r_\mathrm{\ss\mathrm{R}}=0.5\;\mu\mathrm{m}$, $D=D_A=1000\;\mu\mathrm{m}^2/\mathrm{s}$, and $D_B=500\;\mu\mathrm{m}^2/\mathrm{s}$ \cite{noel2015joint}.

In Fig. \ref{comparison}, we plot the MSE of different estimation schemes versus the distance $d$. Here, estimation schemes include the RTT-P \cite{moore2012measuring}, the ML estimation \cite{noel2014bounds}, the peak-based estimation applying a single type of molecules \cite{huang2013distance} and two types of molecules \cite{luo2018effective,sun2017efficient}. We also plot the CRLB as a lower bound to assess the performance of each estimation scheme. For the CRLB and ML estimation, we apply $M=20$ observations. First, we observe that the MSE of ML estimation almost attains the CRLB and hence achieves the best performance. Second, we observe that the peak-based estimation using one type of molecules and RTT-P achieves a moderate performance. The peak-based estimation using one type of molecules outperforms the RTT-P when $d$ is small. When $d$ is large, the advantage of RTT-P becomes more obvious. Third, the peak-based estimation using two types of molecules achieves the worst performance since it calculates the time difference, but does not need synchronization.

\section{Estimation of Other Parameters}\label{pe}

In this section, we review the current studies on the estimation of other parameters (i.e. the parameters excluding distance). We classify these parameters into three categories. The first category is referred to as the \textit{environmental parameters} that are related to channel and $\mathrm{TX}$ properties, e.g., the diffusion coefficient and the number of released molecules. The second category is referred to as the \textit{synchronization-related parameters} that are estimated to achieve synchronization between the $\mathrm{TX}$ and the $\mathrm{RX}$. The third category is referred to as the SNR. We summarize the studies that estimate these parameters in Table \ref{table2}.	

\begin{table*}[!ht]
\newcommand{\tabincell}[2]{\begin{tabular}{@{}#1@{}}#2\end{tabular}}
\centering\caption{Parameter Estimation Schemes for MC Systems}\label{table2}
\begin{tabular}{|c|c|c|c|c|c|c|c|}
\hline
\textbf{Name}&\textbf{\tabincell{c}{Estimated\\ Parameter}}&\textbf{Reference}&\textbf{Method}&\textbf{Noise}&\textbf{TX Waveform}&\textbf{RX Type}&\textbf{Environment}\\
\hline
\multirow{8}{*}{\tabincell{c}{Environmental\\Parameters}}&$d$, $t_0$, $D$,$v$,$k$, $N_\mathrm{tx}$&\cite{noel2015joint}&ML, Peak-Based&RD&Impulse&Transparent&\tabincell{c}{3D, Flow,\\Molecules \\Degradation}\\
\cline{2-8}
&$N_\mathrm{tx}$&\cite{sadeghi2017performance}&HCRLB&RD, ISI&Impulse&Transparent&3D\\
\cline{2-8}
&D&\cite{schafer2019eigenfunction}&Transfer Function&RD&Continuous&\tabincell{c}{Two\\ Fully-absorbing}&1D\\
\cline{2-8}
&$d$, $v$, $k$&\cite{huang2020parameter}&\tabincell{c}{Method of\\Moments}&RD, External&Continuous&\tabincell{c}{Two\\Fully-absorbing}&\tabincell{c}{1D, Flow,\\Molecules\\ Degradation}\\
\cline{2-8}
&$\sigma^2_\mathrm{tx}$&\cite{meng2014receiver}&ML&RD&Impulse&Transparent&1D, 2D, 3D\\
\hline
\multirow{8}{*}{\tabincell{c}{Synchronization\\-Related Parameters}}&\multirow{4}{*}{Clock Offset}&\cite{lin2015diffusion,lin2016clock}&Two-Way, ML&RD&Impulse&Transparent&1D\\
\cline{3-8}
&&\cite{lin2016time}&One-Way, ML&RD&Impulse&Transparent&1D, Flow\\
\cline{3-8}
&&\cite{huang2020clock}&Least Square&RD&\tabincell{c}{Diffusive,\\Impulse}&\tabincell{c}{Diffusive,\\Fully-absorbing}&3D\\
\cline{2-8}
&\multirow{2}{*}{$t_\mathrm{s}(\varepsilon)$}&\cite{jamali2017symbol}&\tabincell{c}{ML, Linear Filter,\\Peak Observation, \\ Threshold-Trigger}&\tabincell{c}{External, ISI,\\RD}&Impulse&Reactive&3D\\
\cline{3-8}
&&\cite{mukherjee2019synchronization}&Faster Molecules&ISI, RD&Impulse&Fully-absorbing&3D\\
\hline
SNR&SNR&\cite{tiwari2016maximum}&ML&ISI, RD&Impulse&Transparent&3D\\
\hline
\end{tabular}
\end{table*}

\subsection{Environmental Parameters}

In this subsection, we focus on the studies that estimate environmental parameters. Specifically, environmental parameters include the distance between the $\mathrm{TX}$ and the $\mathrm{RX}$, release time of molecules, diffusion coefficient of molecules, degradation rate of molecules, flow velocity, and the number of released molecules.

In \cite{noel2015joint}, Noel et al. considered a joint environmental parameter estimation, where the unknown parameter vector $\boldsymbol{\theta}$ contains a single or multiple parameters. By assuming each observation of the received signal at the transparent $\mathrm{RX}$ is independent and follows a Poisson distribution, \cite{noel2015joint} derived the Fisher information matrix $\mathbf{I}(\boldsymbol{\theta})$ and the CRLB when a single unknown parameter exists or multiple unknown parameters exist. Moreover, the ML estimation was applied to estimate the unknown parameters.

In \cite{sadeghi2017performance}, the authors evaluated the HCRLB for a special case via setting $g(\theta)=N_\mathrm{tx}$, i.e., the HCRLB on the variance of the estimated number of released molecules. Similar to \eqref{p_x_theta}, \cite{sadeghi2017performance} derived $p(\mathbf{x};\theta)$ as the joint PMF of the received signals at the transparent $\mathrm{RX}$.

The authors in \cite{noel2015joint,sadeghi2017performance} regarded $N_\mathrm{tx}$ as a constant value in the estimation scheme. Different from that, \cite{meng2014receiver} regarded $N_\mathrm{tx}$ as a RV with the mean of $\mu_\mathrm{tx}$ and variance of $\sigma^2_\mathrm{tx}$. The authors assumed that $\mu_\mathrm{tx}$ is pre-determined and estimated $\sigma^2_\mathrm{tx}$ by emitting multiple impulses of molecules. The $\mathrm{RX}$ detects the received signal at the peak time. By considering that $V_\mathrm{RX}\rightarrow\infty$, i.e., the molecular concentration is perfectly sensed over the entire environment, $\hat{\sigma}^2_\mathrm{tx}$ is obtained via the ML estimation.

It is noted that \cite{noel2015joint,sadeghi2017performance,meng2014receiver} investigated parameter estimation for a single $\mathrm{RX}$ only. Different from these studies,  \cite{schafer2019eigenfunction,huang2020parameter} investigated parameter estimation via two fully-absorbing $\mathrm{RXs}$ in a 1D environment. In \cite{schafer2019eigenfunction}, the authors applied the transfer function to estimate the diffusion coefficient of molecules. Compared to the previous studies, \cite{huang2020parameter} considered the existence of the external additive noise in the parameter estimation process. To reduce the impact of the external additive noise, \cite{huang2020parameter} proposed a novel estimation scheme -- difference estimation (DE) -- to estimate the unknown parameter based on the difference between received signals at two $\mathrm{RXs}$. According to the CIR derived between the $\mathrm{TX}$ and each fully-absorbing $\mathrm{RX}$ in \cite[eq. (8)]{huang2020channel}, \cite{huang2020parameter} derived the CRLB on the variance of the unknown parameter. By assuming that each observation of the received signal is independent and follows a Poisson distribution, $p(\mathbf{x};\theta)$ was obtained as the joint PMF for the difference of received signals at two $\mathrm{RXs}$. In addition, \cite{huang2020parameter} applied the method of moments \cite[Ch.9]{kay1993fundamentals} to estimate the unknown parameter.

\subsection{Synchronization-Related Parameters}

In this subsection, we review the studies on the estimation of synchronization-related parameters that include the clock offset between the $\mathrm{TX}$ and the $\mathrm{RX}$, and the start time of each symbol interval. The clock offset describes a time difference between the $\mathrm{TX}$ and the $\mathrm{RX}$. In a nanonetwork system, different nanomachines work based on their own clocks. Thus, estimating the clock offset is crucial to establish a reliable communication link between the synchronized $\mathrm{TX}$ and $\mathrm{RX}$.

In \cite{lin2015diffusion,lin2016clock}, Lin et al. estimated the clock offset between two transceivers, denoted by $T$ and $R$, respectively, via proposing a two-way message exchange mechanism. In the $\varepsilon$th round of the message exchange, $T$ sends molecules at time $T_{1,\varepsilon}$, and $R$ receives the message at time $T_{2,\varepsilon}$. $R$ then sends a feedback signal at time $T_{3,\varepsilon}$, and $T$ receives the signal at time $T_{4,\varepsilon}$. After $\alpha$ rounds of message exchange, $T$ obtains a set of time instants $\left\{T_{1,\varepsilon}, T_{2,\varepsilon}, T_{3,\varepsilon}, T_{4,\varepsilon}\right\}_{\varepsilon=1}^{\alpha}$. By assuming that the propagation delay follows an inverse Gaussian distribution and Gaussian distribution in \cite{lin2015diffusion} and \cite{lin2016clock}, respectively, the joint PDF of the molecular propagation delay for the $\alpha$-round message exchange can be obtained. Based on the PDF, the clock offset is estimated via the ML estimation. After that, $R$ can be synchronized to $T$. Moreover, \cite{lin2016time} proposed a one-way clock offset estimation method due to the fact that two-way estimation has a high demand for transceivers and is time-consuming as aforementioned in Section \ref{MD}. Based on the joint PDF of the molecular propagation delay for multiple transmissions from the $\mathrm{TX}$, the clock offset is estimated by the ML estimation. Furthermore, \cite{huang2020clock} considered clock offset estimation when both $\mathrm{TX}$ and $\mathrm{RX}$ are diffusive mobile. After molecules are released from the $\mathrm{TX}$, the $\mathrm{RX}$ counts the number of arrived molecules for $M$ times. The authors estimated clock offset by using the least square method that finds the clock offset to minimize the sum of differences between the mean of the CIR over the varying distance and $M$ observations.

The estimation of clock offset is adequate to achieve synchronization only if the clock offset is fixed. To overcome this issue, \cite{jamali2017symbol,mukherjee2019synchronization} considered the estimation of the start time of the $\eta$th symbol interval, denoted by $t_\mathrm{s}[\eta]$. In \cite{jamali2017symbol}, the authors first proposed the ML estimation scheme to estimate $t_\mathrm{s}[\eta]$. By considering that each observation within this interval follows a Poisson distribution, $t_\mathrm{s}[\eta]$ is estimated based on the joint PDF of multiple observations. Due to the high complexity of the ML estimation, the authors then proposed three suboptimal low-complexity estimation schemes. The first suboptimal estimation scheme is a linear filter-based scheme that finds $t_\mathrm{s}[\eta]$ to maximize the expected mean of multiple observations. The second one is the peak observation-based scheme that estimates $t_\mathrm{s}[\eta]$ based on the peak observation at the $\mathrm{RX}$. The third one is the threshold-trigger scheme that determines $t_\mathrm{s}[\eta]$ when the observation is larger than a predefined threshold. Notably, \cite{jamali2017symbol} considered the impact of the external additive noise and ISI on these estimation schemes. Moreover, \cite{mukherjee2019synchronization} considered the transmission of molecules with a faster diffusion coefficient than information molecules to realize synchronization, where $t_\mathrm{s}[\eta]$ is estimated as the time when the peak concentration of faster molecules is detected at the $\mathrm{RX}$. 
\subsection{SNR}

In \cite{tiwari2016maximum}, Tiwari et al. estimated the SNR in MC system, which considered the noise induced by the ISI. In \cite{tiwari2016maximum}, the SNR was defined as
\begin{align}
\mathrm{SNR}=\frac{P_\mathrm{s}}{P_\mathrm{n}},
\end{align}
where $P_\mathrm{s}$ represents the power of the intended received signal at the transparent $\mathrm{RX}$ and $P_\mathrm{n}$ represents the power of noise due to the ISI. The power in MC can be interpreted as the square number of molecules. Thus, the $\mathrm{SNR}$ is a function of the number of emitted molecules $N_\mathrm{tx}$ and the variance of the noise, denoted by $\sigma^2_\mathrm{n}$, where the noise is regarded as a Gaussian RV. According to the invariance property of the ML estimation, estimating a function with multiple unknown parameters is equivalent to estimating individual unknown parameters \cite{kay1993fundamentals}. Therefore, the SNR is estimated by using the ML estimation of $N_\mathrm{tx}$ and $\sigma^2_\mathrm{n}$ based on the joint PDF of the received signals at the $\mathrm{RX}$. In addition, the CRLB was derived when $\boldsymbol{\theta}=\left[N_\mathrm{tx}^2, \sigma^2_\mathrm{n}\right]$.
	
\section{Channel Estimation}\label{ce}

\begin{table*}[!ht]
\newcommand{\tabincell}[2]{\begin{tabular}{@{}#1@{}}#2\end{tabular}}
\centering
\caption{CIR Estimation Schemes for MC Systems}
\label{tab:tab3}
\begin{tabular}{|c|c|c|c|c|c|c|}
\hline
\textbf{Name}                                                                                   & \textbf{References}                                    & \textbf{Method}                                                                     & \textbf{Performance}                                                                   & \textbf{Computational Complexity}                                                      & \textbf{Noise}                                                                      & \textbf{TX Waveform} \\ \hline
\multirow{3}{*}{\begin{tabular}[c]{@{}c@{}}Pilot-Based \\ CIR Estimation\end{tabular}} & \cite{7546910}                                & ML                                                                         & \multirow{5}{*}{\begin{tabular}[c]{@{}c@{}}EM\textgreater{}DD\textgreater{}ML\\ \textgreater{}LSSE(LO)\end{tabular}} & \multirow{5}{*}{\begin{tabular}[c]{@{}c@{}}EM\textless{}DD\textless{}ML\\ \textless{}LSSE(LO)\end{tabular}} & RD,ISI                                                                     & Impulse     \\ \cline{2-3} \cline{6-7} 
                                                                                       & \cite{7546910},\cite{wang2020understanding} & LSSE                                                                       &                                                                               &                                                                               & RD, ISI                                                                    & Impulse     \\ \cline{2-3} \cline{6-7} 
                                                                                       & \cite{8685177}                                & ML,LSSE                                                                    &                                                                               &                                                                               & \begin{tabular}[c]{@{}c@{}}RD,ISI, ILI\end{tabular} & Impulse     \\ \cline{1-3} \cline{6-7} 
\multirow{2}{*}{\begin{tabular}[c]{@{}c@{}}Semi-Blind\\  CIR Estimation\end{tabular}}  & \cite{9146169}                                & EM,DD                                                                      &                                                                               &                                                                               & RD, ISI                                                                    & Impulse     \\ \cline{2-3} \cline{6-7} 
                                                                                       & \cite{9340139}                                & LO                                                                         &                                                                               &                                                                               & RD,ISI                                                                     & Impulse     \\ \hline
\begin{tabular}[c]{@{}c@{}}CIR Estimation \\ Used at RX Design\end{tabular}            & \cite{6708551}                                & \begin{tabular}[c]{@{}c@{}}MSE, \\ Steepest-Descent Algorithm\end{tabular} & N/A                                                                           & N/A                                                                           & RD,ISI                                                                     & Impulse     \\ \hline
\end{tabular}
\end{table*}
In this section, we first present the CIR estimation problem and then review different CIR estimation schemes. We also summarize different CIR estimation studies in Table \ref{tab:tab3}. 

\subsection{Problem Formulation}

We consider an MC system as shown in Fig. \ref{sys3d}. At the beginning of each symbol interval, the $\mathrm{TX}$ releases $N_\mathrm{tx}$ molecules if the transmitted symbol is ``1'', but does not release any molecule if the transmitted symbol is ``0''. Taking into account the effect of ISI, we assume the input-output relationship of the MC system as
\begin{align}\label{Eq:ChannelInOut}
z[q]=\sum_{u=1}^{U} c_u[q] + c_{\mathtt{n}}[q],
\end{align}
where $z[q]$ is the number of molecules detected at the $\mathrm{RX}$ in symbol interval $q$, $U$ is the number of memory taps of the channel, and $c_u[q]$ is the number of molecules observed at the $\mathrm{RX}$ in symbol interval $u$, due to the release of $b[q-u+1]N_\mathrm{tx}$ molecules by the $\mathrm{TX}$ in symbol interval $q-u+1$, where $b[q]\in[0,1]$ is the transmitted symbol in symbol interval $q$. Therefore, $c_u[q]$ can be well approximated by a Poisson RV with the mean of $\overline{c}_u b[q-u+1]$. Moreover, $c_{\mathtt{n}}[q]$ is the number of external additive noise molecules detected by the $\mathrm{RX}$ in the symbol interval $q$. Let $\mathbf{b}=[b[1],b[2],\dots,b[Q]]^T$ be a training sequence of length $Q$.

For convenience of notation, we define $\mathbf{z}=[z[U],z[U+1],\dots,z[Q]]^T$, $\overline{\mathbf{c}}=[\overline{c}_1, \overline{c}_2,\dots,\overline{c}_U,\overline{c}_{\mathtt{n}}]^T$ is the CIR of the channel, and $f_{\mathbf{z}}(\mathbf{z}|\overline{\mathbf{c}},\mathbf{b})$ is the PDF of the observation $\mathbf{z}$ conditioned on a given channel $\overline{\mathbf{c}}$ and a given training sequence $\mathbf{b}$. The goal of channel estimation is to estimate $\overline{\mathbf{c}}$ based on the vector of random observations $\mathbf{z}$.

\subsection{Pilot-Based CIR Estimation}

In this subsection, we present the pilot-based CIR estimation scheme studied in \cite{7546910}\footnote{We clarify that both [15] and [16] are technical papers rather than comprehensive surveys on MC channel estimation.}, where the transmission of a known training sequence of pilots is required for the estimation and calculation of the corresponding CRLB.

\subsubsection{ML Estimation}

The ML CIR estimation scheme aims to find the CIR that maximizes the likelihood of observation vector $\mathbf{z}$ \cite{BayesianBook}. In particular, the ML estimation is given by
\begin{align}\label{Eq:ML_Estimation}
\hat{\overline{\mathbf{c}}}^{\mathtt{ML}} = \underset{\overline{\mathbf{c}}\geq \mathbf{0}}{\mathrm{argmax}} \,\,f_{\mathbf{z}}(\mathbf{z}|\overline{\mathbf{c}},\mathbf{b}).
\end{align}
We assume that $z[q]$ is a Poisson RV with the mean of $\overline{z}[q] = \overline{c}_{\mathtt{n}}+\sum_{u=1}^{U} \overline{c}_u b[q-u+1] = \overline{\mathbf{c}}^T\mathbf{b}_q$ and $\mathbf{b}_q=[b[q],b[q-1],\dots,b[q-U+1],1]^T$. Under this assumption, $f_{\mathbf{z}}(\mathbf{z}|\overline{\mathbf{c}},\mathbf{b})$ is given by
\begin{align}\label{Eq:ML_PDF}
f_{\mathbf{z}}(\mathbf{z}|\overline{\mathbf{c}},\mathbf{b})\,\,
=\prod_{q=U}^{Q}\frac{\left(\overline{\mathbf{c}}^T\mathbf{b}_q\right)^{z[q]}}{z[q]!}
\exp\left(-\overline{\mathbf{c}}^T\mathbf{b}_q\right).
\end{align}

We note that \cite{7546910} solved the ML estimation of the CIR given in \eqref{Eq:ML_Estimation} by using \textbf{Algorithm 1}, where the following non-linear system of equations is solved\footnote{A system of nonlinear equations can be solved by using mathematical software packages, e.g., Mathematica.} for different $\mathcal{A}_w$
\begin{align}\label{Eq:ML_Sol}
\sum_{q=L}^{Q}\left[\frac{z[q]}{(\overline{\mathbf{c}}^{\mathcal{A}_{w}})^T \mathbf{b}_q^{\mathcal{A}_w}}-1\right]\mathbf{b}_q^{\mathcal{A}_w}=\mathbf{0},
\end{align}
where $\mathcal{A}=\{\mathcal{A}_1,\mathcal{A}_2,\cdots,\mathcal{A}_W\}$ denotes a set which contains all possible $W=2^{U+1}-1$ subsets of the set $\mathcal{F}=\{1,2,\cdots,U,\mathtt{n}\}$, except for the empty set. Here, $\mathcal{A}_w$ denotes the $w$-th subset of $\mathcal{A}$, $w=1,2,\cdots,W$. Moreover, let $\overline{\mathbf{c}}^{\mathcal{A}_w}$ and $\mathbf{b}_q^{\mathcal{A}_w}$ denote the reduced-dimension versions of $\overline{\mathbf{c}}$ and $\mathbf{b}_q$, respectively, which only contain those elements of $\overline{\mathbf{c}}$ and $\mathbf{b}_q$ whose indices are the elements of $\mathcal{A}_w$, respectively.

\begin{algorithm}[t]\label{alg_1}
\caption{{ML} (or {LSSE}) CIR estimation of {$\hat{\overline{\mathbf{c}}}^{\mathtt{ML}}$} (or {$\hat{\overline{\mathbf{c}}}^{\mathtt{LSSE}}$)} \cite{7546910}		}
\begin{algorithmic}
\State \textbf{initialize} $\mathcal{A}_w=\mathcal{F}$ and solve {(\ref{Eq:ML_Sol})} (or {(\ref{Eq:LSSE_Sol})}) to find $\overline{\mathbf{c}}^{\mathcal{F}}$
\If{$\overline{\mathbf{c}}^{\mathcal{F}}\geq \mathbf{0}$}
\State  Set {$\hat{\overline{\mathbf{c}}}^{\mathtt{ML}} = \overline{\mathbf{c}}^{\mathcal{F}}$} (or {$\hat{\overline{\mathbf{c}}}^{\mathtt{LSSE}} = \overline{\mathbf{c}}^{\mathcal{F}}$})
\Else
\For{$\forall \mathcal{A}_w\neq \mathcal{F}$}
\State Solve (\ref{Eq:ML_Sol}) (or (\ref{Eq:LSSE_Sol})) to find $\overline{\mathbf{c}}^{\mathcal{A}_w}$
\If{$\overline{\mathbf{c}}^{\mathcal{A}_w}\geq\mathbf{0}$ holds}
\State Set the values of the elements of $\hat{\overline{\mathbf{c}}}^{\mathtt{CAN}}$, whose indices are in $\mathcal{A}_w$, equal to the values of the corresponding elements in $\overline{\mathbf{c}}^{\mathcal{A}_w}$ and the remaining $U+1-\left|\mathcal{A}_w\right|$ elements equal to zero;
\State Save $\hat{\overline{\mathbf{c}}}^{\mathtt{CAN}}$ in the candidate set $\mathcal{C}$
\Else
\State Discard $\overline{\mathbf{c}}^{\mathcal{A}_w}$
\EndIf
\EndFor
\State Choose  $\hat{\overline{\mathbf{c}}}^{\mathtt{ML}}$ (or $\hat{\overline{\mathbf{c}}}^{\mathtt{LSSE}}$) equal to $\hat{\overline{\mathbf{c}}}^{\mathtt{CAN}}$ in the candidate set $\mathcal{C}$ which  maximizes $g(\overline{\mathbf{c}})$ (or minimizes $\|\boldsymbol{\epsilon}\|^2$).
\EndIf
\end{algorithmic}
\end{algorithm}

\subsubsection{Least Sum of Squared Errors (LSSE) CIR Estimation}

The LSSE CIR estimation scheme aims to choose $\overline{\mathbf{c}}$ that minimizes the sum of the squared errors for the observation vector $\mathbf{z}$. Here, the error vector is defined as $\boldsymbol{\epsilon} = \mathbf{z} - \E\left\{\mathbf{z}\right\} = \mathbf{z} - \mathbf{B} \overline{\mathbf{c}}$ where $\mathbf{B} = [\mathbf{b}_U,\mathbf{b}_{U+1},\dots,\mathbf{b}_Q]^T$. In particular, the LSSE CIR estimation can be written as
\begin{align}\label{Eq:LSSE_Estimation}
\hat{\overline{\mathbf{c}}}^{\mathtt{LSSE}}=\underset{\overline{\mathbf{c}}\geq \mathbf{0}}{\mathrm{argmin}} \,\, \| \boldsymbol{\epsilon} \|^2.
\end{align}

It is noted that \cite{7546910} also obtained the LSSE estimation of the CIR by solving \eqref{Eq:LSSE_Estimation}. The solution is given by \textbf{Algorithm 1} where for a given set $\mathcal{A}_w$, $\overline{\mathbf{c}}^{\mathcal{A}_w}$ is obtained as
\begin{align}\label{Eq:LSSE_Sol}
\overline{\mathbf{c}}^{\mathcal{A}_w}=\left((\mathbf{B}^{\mathcal{A}_w})^T \mathbf{B}^{\mathcal{A}_w}\right)^{-1}(\mathbf{B}^{\mathcal{A}_w})^T\mathbf{z}.
\end{align}

We also note that \cite{wang2020understanding} designed a communication protocol that estimates the CIR based on the least square method, similar to the LSSE CIR estimation in \cite{7546910}.

\subsubsection{CRLB}

With the estimation error vector defined as $\mathbf{e} = \overline{\mathbf{c}} - \hat{\overline{\mathbf{c}}}$, the classical CRLB for the deterministic $\overline{\mathbf{c}}$ provides the following lower bound on the sum of the expected square errors \cite{7546910}
\begin{align}\label{Eq:CRB_CIR}
\E\left\{\|\mathbf{e}\|^2\right\}
\geq\mathrm{tr}\left\{\mathbf{I}^{-1}\left(\overline{\mathbf{c}}\right)\right\}
=\mathrm{tr}\left\{\left[\sum_{q=U}^{Q}\frac{\mathbf{b}_q\mathbf{b}_q^T}
{\overline{\mathbf{c}}^T\mathbf{b}_q}\right]^{-1}\right\},
\end{align}
where $\mathrm{tr}\{\cdot\}$ denotes the trace of a matrix. While \cite{7546910} derived the ML and LSSE estimation schemes of the CIR and the CRLB for a single-input single-output channel, \cite{8685177} extended these results to a diffusive MIMO system, by incorporating the ILI.

\subsection{Semi-Blind CIR Estimation}

In this subsection, we present semi-blind CIR estimation schemes where the transmission of $Q$ pilot symbols is followed by the transmission of $\mathcal{D}$ unknown data symbols, denoted by the vector $\boldsymbol{\beta}=[\beta[1],\beta[2],\dots,\beta[\mathcal{D}]]^T$. The semi-blind CIR estimation schemes incorporate both data-carrying and pilot-carrying observations into the estimation process. This is different from pilot-based estimation which only considers the received pilot-carrying observations in the estimation process but excludes data-carrying observations. Incorporating data-carrying observations into the estimation can significantly enhance the estimation accuracy and/or improve the data rate \cite{9146169}.

An expectation maximization (EM)-based estimation scheme was first proposed in \cite{9146169}. The data vector $\boldsymbol{\beta}$ constitutes hidden information at the $\mathrm{RX}$. Beginning with the initial guess, the EM estimation scheme alternates between obtaining the conditional expectation of the complete-data log-likelihood and maximizing the result with respect to the desired parameters. The $\omega$th iteration of the EM estimation scheme consists of two steps. The first is the expectation step (E-step) which consists of obtaining the expectation of the complete-data log-likelihood function. This is followed by the maximization step (M-step), in which the \emph{posterior} probability of the hidden data is maximized to acquire an updated estimate of the CIR. Abdallan et al. in \cite{9146169} also proposed two semi-blind estimation schemes based on the decision-directed (DD) strategy. The idea of the DD strategy is to use the channel estimate acquired through pilot-based estimation for performing data detection. The detected data is in turn treated as a new set of pilots to perform another cycle of channel estimation \cite{9146169}. We note that the analytical derivation of the CRLB can be very challenging in semi-blind estimation, because the log-likelihood function of observations becomes complicated when the statistics of data symbols are taken into account. In \cite{9146169}, the authors applied the Monte-Carlo method to obtain accurate approximations of the Fisher information matrix $\mathbf{I}^{-1}\left(\bar{\mathbf{c}}\right)$. The semi-blind CRLB was then obtained by evaluating $\mathrm{tr}\left\{\mathbf{I}^{-1} \left(\bar{\mathbf{c}}\right)\right\}$.

Darya et al. in \cite{9340139} proposed a modified version of the decision-directed least-squares (DDLS) estimation scheme proposed in \cite{9146169}. In \cite{9340139}, the authors named the proposed estimation scheme as the low-overhead (LO) estimation scheme which has reduced complexity compared to the DDLS estimation scheme. The authors in \cite{9340139} showed that the LO
estimation scheme has comparable performance to the pilot-based LS estimation scheme in \cite{7546910}, while achieving a reduction of up to 95\% in the pilot overhead by using a minimal number of pilot symbols.

\subsection{Pilot-Based Estimation versus Semi-Blind Estimation}
\begin{figure}[!t]
    \begin{center}
	\includegraphics[height=2in]{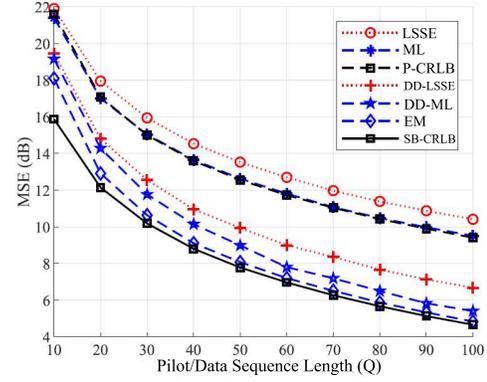}
    \caption{The MSE of different CIR estimation schemes versus pilot sequence length \cite[Fig. 1]{9146169}. LSSE: least squares, ML: maximum-likelihood,  P-CRLB: Pilot-Based CRLB, DD-ML: decision-directed ML, DD-LS: decision-directed LS, SB-CRLB: semi-blind CRLB.}\label{comparison1}\vspace{-0.5em}
	\end{center}\vspace{-4mm}
\end{figure}

Based on the simulation results in Fig. \ref{comparison1} (we re-presented Fig. 1 in \cite{9146169} as Fig. \ref{comparison1} here for convenience), the semi-blind estimation schemes achieve a significantly lower MSE than the existing pilot-based ML and LSSE estimation schemes. Also, the semi-blind estimation schemes can substantially reduce the pilot overhead as compared to the best-performing pilot-based estimation schemes, by more than 60\% for the case of EM and 55\%--44\% for DD-based estimation. The EM-based semi-blind estimation scheme provides the highest estimation accuracy. The DD-based semi-blind estimation scheme performs almost midway between the pilot-based ML and the EM-based semi-blind estimation schemes, but achieves a lower computational cost than the EM-based semi-blind estimation scheme.

\subsection{Channel Estimation Scheme at the RX design}
In this subsection, we present a joint channel and data estimation scheme that was presented in \cite{6708551}, while \cite{7546910,8685177,9146169} only focused on channel estimation based on given training sequences. Specifically, \cite{6708551} presented an estimation scheme which is compatible with different detectors (e.g., the MAP  and ML sequence detectors) at the RX to jointly recover the transmitted bits from the molecule observations distorted by both the ISI and noise. This is because that the detectors at the RX require the knowledge of the CIR. 
In \cite{6708551}, the channel estimation scheme uses a steepest-descent algorithm to recursively estimate the CIR which minimizes the MSE between the actual received sequence and the output of the estimation scheme. The speed of the convergence and the accuracy of the estimation are determined by the value of the step size in the steepest-descent algorithm. Also, due to the use of the steepest-descent algorithm, the channel estimation scheme is able to track slow variations in the CIR.

\section{Future Research Directions}\label{frd}

In this section, we identify and discuss some future research directions for parameter estimation and channel estimation in MC systems. Current studies on estimation have usually considered simple $\mathrm{TX}$ and $\mathrm{RX}$ protocols and ideal communication channels. These simplifications would lead to inaccuracy when the current estimation schemes are applied into practical MC environments. Moreover, the noise is common in MC systems. Effective methods to mitigate the impact of noise on the estimation performance need to be widely investigated. Based on these, we present some open research problems as follows:
\begin{itemize}
\item \textbf{Imperfect TX:} Current studies on estimation schemes have considered an ideal point $\mathrm{TX}$. Compared to realistic scenarios, this ideal $\mathrm{TX}$ model does not address the properties of the $\mathrm{TX}$, such as geometry, signaling pathways inside the $\mathrm{TX}$, and chemical reactions during the release process. Some recent studies have proposed imperfect $\mathrm{TX}$ models, e.g., an ion channel-based $\mathrm{TX}$ in \cite{arjmandi2016ion} and a membrane fusion-based $\mathrm{TX}$ in \cite{huang2020membrane}. Applying these imperfect $\mathrm{TX}$ into estimation schemes is an interesting future work.
\item \textbf{Macro-scale estimation:} Most current studies have examined the estimation in the micro-scale environment while a few studies have considered macro-scale estimation. As aforementioned, MC also exists in macro-scale. Thus, the estimation in the macro-scale environment needs attention. Here, we propose two research directions. The first direction is to perform estimation based on tabletop experiments. Some studies have established the tabletop experiment for macro-scale MC, e.g., \cite{farsad2013tabletop,mcguiness2019experimental,grebenstein2019molecular}. Estimation can be investigated based on these experiments. The second direction is the theoretical analysis of estimation schemes in practical macro-scale environments, e.g., a pipe or river. Compared to traditional theoretical analysis in micro-scale MC, flow modeling is crucial for the analysis in the macro-scale environment. For example, the flow can be modeled as laminar in the pipe and turbulent in the river.
\item \textbf{Noise mitigation:} Most studies have included noise, e.g., RD, ISI, and external additive noise, in the estimation process, while only a few of them, e.g., \cite{huang2020parameter}, proposed methods to mitigate the impact of noise on estimation. It is noted that \cite{huang2020parameter} only focused on a stable stage of the communication channel, i.e., the expected received signal is constant, when time is large. More general noise mitigation methods should be investigated during the estimation process.
\item \textbf{MIMO estimation:} Most of existing studies have focused on the estimation via one TX and one RX. Only a limited number of studies, such as \cite{huang2020parameter}, investigated cooperative estimation via two RXs and showed the improved estimation accuracy as compared to the single-RX estimation. Meanwhile, a few studies, such as \cite{8685182}, showed that the detection performance of transmitted symbols is greatly improved by combining the received information at multiple distributed receivers. Based on these studies, the performance enhancement in parameter or channel estimation by combining the received information at multiple receivers has not been thoroughly explored. Moreover, estimation schemes by using the MIMO system have not been investigated. In particular, multiple channels can be adopted to estimate multiple unknown parameters.
\item \textbf{Real-time parameter estimation:} The existing studies have focused on one-shot parameter estimation by assuming that the estimated parameters are constant over time, or have focused on the estimation of the initial value of parameters when the estimated parameters keep changing over time. However, in dynamic biological environments, many parameters vary over time, e.g., the distance between the TX and RX changes when the transceivers move. Hence, the real-time estimation of time-varying parameters is a promising research direction.		
\end{itemize}

\section{Conclusion}\label{c}
	
In this paper, we provided the first comprehensive survey on parameter estimation and channel estimation in MC systems. We summarized three types of noise that can influence the estimation performance and three metrics that can be used to evaluate the performance of an estimation scheme. For parameter estimation, we first presented a detailed review on the distance estimation and compare the performance of different estimation schemes via calculating the MSE. We then provided a detailed review on the estimation of other parameters. For channel estimation, we reviewed the pilot-based CIR estimation scheme and the semi-blind CIR estimation scheme. Finally, we discussed some open research problems for parameter estimation and channel estimation. This survey helps MC researchers to develop an in-depth understanding on the current estimation schemes in MC and serves as a cornerstone for MC researchers to explore more advanced estimation schemes in the future.
\bibliographystyle{IEEEtran}
\bibliography{references}
\end{document}